\documentclass[aps,prm,twocolumn,superscriptaddress,longbibliography]{revtex4-2}
\usepackage{amsmath,amssymb}
\usepackage[pdftex]{hyperref,graphicx}
\hypersetup{colorlinks = true, urlcolor = blue, linkcolor = blue, citecolor = blue}
\usepackage{physics}
\usepackage{xcolor}
\usepackage{bm}
\usepackage{pdfpages}

\newcommand{\comment}[1]{{}}

\makeatletter
\AtBeginDocument{\let\LS@rot\@undefined}
\makeatother

\begin{document}
\title{Density-dependent two-dimensional optimal mobility in ultra-high-quality semiconductor quantum wells}
\author{Seongjin Ahn}
\affiliation{Condensed Matter Theory Center and Joint Quantum Institute, Department of Physics, University of Maryland, College Park, Maryland 20742, USA}
\author{Sankar Das Sarma}
\affiliation{Condensed Matter Theory Center and Joint Quantum Institute, Department of Physics, University of Maryland, College Park, Maryland 20742, USA}

\begin{abstract}
We calculate using the Boltzmann transport theory the density dependent mobility of two-dimensional (2D) electrons in GaAs, SiGe and AlAs quantum wells as well as of 2D holes in GaAs quantum wells.  The goal is to precisely understand the recently reported breakthrough in achieving a record 2D mobility for electrons confined in a GaAs quantum well. Comparing our theory with the experimentally reported electron mobility in GaAs quantum wells, we conclude that the mobility is limited by unintentional background random charged impurities at an unprecedented low concentration of $\sim10^{13} \mathrm{cm}^{-3}$.  We find that this same low level of background disorder should lead to 2D GaAs hole and 2D AlAs electron mobilities of $\sim10^7 \mathrm{cm}^2/Vs$ and $\sim4\times10^7 \mathrm{cm}^2/Vs$, respectively, which are much higher theoretical limits than the currently achieved experimental values in these systems. We therefore conclude that the current GaAs hole and AlAs electron systems are much dirtier than the state of the arts 2D GaAs electron systems. We present theoretical results for 2D mobility as a function of density, effective mass, quantum well width, and valley degeneracy, comparing with experimental data. 

\end{abstract}

\maketitle
\section{Introduction and Background}
Modulation doping \cite{dingleElectronMobilitiesModulation1978}
initiated the modern era of semiconductor-based high mobility two-dimensional (2D) carrier systems simply by spatially separating the dopant atoms from the carriers released by the dopants, thus suppressing the detrimental effect of impurity scattering on carrier transport. During the first 30 year period of 1978-2008, the 2D mobility in the archetypal $n$-GaAs-based 2D electron system increased by more than a factor $1000$, from $2\times10^4 \mathrm{cm}^2/Vs$ in 1978 to $3\times10^7 \mathrm{cm}^2/Vs$ in 2008, keeping pace with the famous Moore's Law in microelectronics, through materials improvement in the molecular beam epitaxy (MBE) technique used in producing high-quality semiconductor quantum wells hosting the 2D confined carriers \cite{chungUltrahighqualityTwodimensionalElectron2021, chungWorkingPrinciplesDopingwell2020}. 
This is an astonishing materials physics accomplishment, which led to a revolution in fundamental experimental condensed matter physics, leading to the laboratory observations of the fractional quantum Hall effect \cite{tsuiTwoDimensionalMagnetotransportExtreme1982},
the eve$n$-denominator fractional quantum Hall effect \cite{willettObservationEvendenominatorQuantum1987},
bilayer fractional quantum Hall effects 
\cite{suenObservationFractionalQuantum1992, eisensteinNewFractionalQuantum1992}, 
Wigner crystallization \cite{yoonWignerCrystallizationMetalInsulator1999},
and many other phenomena far too numerous to cite here. Unfortunately, this whole development came to an abrupt halt in 2008 with no further improvement in the 2D mobility during the 2008-2020 period in spite of concerted efforts by several MBE groups \cite{chungUltrahighqualityTwodimensionalElectron2021}. 

Very recently, however, there has been a breakthrough in the MBE growth of 2D GaAs-AlGaAS quantum wells leading to a sudden abrupt increase in the mobility to $44\times10^6 \mathrm{cm}^2/Vs$, with improvement to even higher mobilities very likely in the near future. This recent breakthrough arises from the rather mundane effect of improving the basic semiconductor quality during growth so that the unintentional background charged impurity concentration in the system is brought down to an incredibly low number of $10^{13} \mathrm{cm}^{-3}$ as described in depth in Ref.~\cite{chungUltrahighqualityTwodimensionalElectron2021}. 
A further decrease in the background impurity density under even cleaner MBE growth conditions may soon lead to the goal of achieving the `100-million mobility' in 2D systems \cite{hwangLimitTwodimensionalMobility2008}.
The background unintentional doping, rather than modulation doping (or interface roughness scattering), is known to be the mechanism limiting the low-temperature mobility in GaAs-AlGaAs based 2D electron systems, because the modulation doping layer is simply too far spatially to cause significant resistive scattering (although it may still control some aspects of the `quality' \cite{dassarmaMobilityQualityTwodimensional2014a})  
and because the layer-by-layer nature of MBE growth assures high quality epitaxial interfaces. In fact, the behavior of the 2D mobility as a function of the carrier density is a sharp diagnostic for the nature of the limiting low temperature scattering mechanism in the 2D system \cite{dassarmaUniversalDensityScaling2013, dassarmaTransportTwodimensionalModulationdoped2015a} 
as has been known for a long time \cite{shayeganGrowthLowDensity1988}, 
and all MBE-grown high-mobility 2D carrier systems are known to be limited by background impurity scattering for more than 30 years.  Therefore, the finding in Ref.~\cite{chungUltrahighqualityTwodimensionalElectron2021} 
that improving the materials quality leads to higher mobility is expected, but is nevertheless an important experimental achievement. To emphasize the importance of this breakthrough, we mention that the existing 2D $n$-GaAs mobility record is $35\times10^6 \mathrm{cm}^2/Vs$ for a density of $3\times10^{11} \mathrm{cm}^{-2}$ \cite{umanskyMBEGrowthUltralow2009}, 
which corresponds only to a mobility of $\sim25\times10^6 \mathrm{cm}^2/Vs$ at the density $\sim10^{11} \mathrm{cm}^\mathrm{-2}$ where the record mobility of Ref.~\cite{chungUltrahighqualityTwodimensionalElectron2021} 
is reported (assuming everything else remains the same). Thus, the new record mobility is almost a factor of 2 improvement in the background disorder content over the existing situation!

In the current work, we analyze in depth the reported 2D $n$-GaAs mobility results in Ref.~\cite{chungUltrahighqualityTwodimensionalElectron2021} 
using a realistic transport theory, obtaining the magnitude of the background random impurity density by comparing our theory with the data.  This enables us to predict how the low-temperature mobility should improve with increasing (decreasing) carrier (impurity) density in the future.  We also calculate the predicted 2D mobility in equivalently clean 2D $p$-GaAs, $n$-AlAs, and $n$-SiGe modulatio$n$-doped quantum-well systems, finding that the current state of the arts experimental mobilities in these other 2D systems are much lower than the theoretical predictions, implying that the MBE growth of these systems is still much dirtier than that achieved in Ref.~\cite{chungUltrahighqualityTwodimensionalElectron2021} 
for electrons in 2D GaAs-AlGaAs quantum well structures.  We provide theoretical results as functions of carrier density, impurity density, quantum well width, effective mass, and valley degeneracy for completeness and future reference.

\section{Theory and Results}

The only resistive scattering mechanism we consider is scattering by background unintentional random charged impurities, which are invariably present in all materials.  Other scattering mechanisms, such as scattering by remote dopants in the modulation doping layer or by interface roughness or by phonons, etc., are miniscule for the experimental and theoretical situations of our interest. We use the well-known Boltzmann transport theory and screened Coulomb disorder \cite{andoElectronicPropertiesTwodimensional1982}. 
The low-temperature mobility is given by:
\begin{equation} \label{eq:mobility}
\mu = e\tau_\mathrm{F}/m,    
\end{equation}
where $m$ is the 2D carrier effective mass, and $\tau_\mathrm{F}$ is the transport relaxation time at the Fermi surface in the Boltzmann theory which we calculate in the leading order scattering approximation for the screened charged impurity potential:
\begin{equation} \label{eq:transport_relaxation_time}
    \frac{1}{\tau_\mathrm{F}}=\frac{2\pi}{\hbar} \int N_i(z) dz 
    \sum_{\bm k'}
    \left|V_{\bm k_\mathrm{F} - \bm k'}(z)\right|^2
    (1-\cos{ \theta})\delta(\epsilon_{\bm k_\mathrm{F}}-\epsilon_{\bm k'}),
\end{equation}
where $\epsilon_{\bm k}=\hbar^2k^2/2m$ is the usual parabolic energy dispersion, $N_i(z)$ is the three dimensional distribution of impurities with $z$ being the distance from the center of the quantum well, $\theta$ is the scattering angle between $\bm k$ and $\bm k'$, and $V_{\bm q}(z)=(v^{(c)}_{\bm q}/\varepsilon_{\bm q}) e^{-q\left|z\right|}F^{(i)}_{\bm q}$ is the electro$n$-impurity scattering matrix element. Here $v_{\bm q}^{(c)}=2\pi e^2/\kappa q$ is the Coulomb interaction with $\kappa$ representing the dielectric constant, and $\varepsilon_{\bm q}=1+v^{(c)}_{\bm q} F_{\bm q} \Pi_{\bm q}$ is the static screening function where $\Pi_{\bm q}$ is the noninteracting polarization function given by \cite{sternPolarizabilityTwoDimensionalElectron1967}
 \begin{equation}
    \Pi_{\bm q}=
		\frac{g_\mathrm{v}m}{\pi\hbar^2}\left[1 - \Theta(q-2k_\mathrm{F})\frac{\sqrt{q^2- 4k^2_\mathrm{F} }}{q}\right],
    \label{eq:diel}
\end{equation}
and $g_\mathrm{v}$ denotes the valley degeneracy.
For realistic calculations, we include the quantum well form factors $F^{(i)}_{\bm q}$ and $F_{\bm q}$ in our calculations to take into account the effects of quantum well thickness, which are given by \cite{hwangElectronicTransportTwodimensional2013a}
\begin{equation}
    F_{\bm q}=\frac{ 3(qa) + 8\pi^2/(qa)}{(qa)^2 + 4\pi^2} - \frac{32\pi^4(1-e^{-qa})}{(qa)^2[(qa)^2 + 4\pi^2 ]^2},
\end{equation}
and
\begin{equation} \label{eq:5}
    F^{(i)}_{\bm q}= \frac{4}{qa} \frac{ 2\pi^2(1-e^{-qa/2}) + (qa)^2 }{ (4\pi)^2 + (qa)^2 },
\end{equation}
where $a$ is the quantum well width. 

We note that the 2D mobility, as defined above, depends on several parameters: $n$, $n_\mathrm{i}$, $m$, $a$, $g_v$, $\kappa$. For a given system, $m$, $a$, $g_v$, and $\kappa$ are fixed and known, with the 2D carrier density $n$ being the only experimentally tunable sample-dependent known parameter. The background charged impurity density $n_\mathrm{i}$ is also a variable, but it is unknown by definition since it varies randomly from sample to sample. We therefore vary $n$ and $n_\mathrm{i}$ to obtain our mobility results, fitting the theory to the data presented in \cite{chungUltrahighqualityTwodimensionalElectron2021} 
by varying $n$ (known from experiment) and $n_\mathrm{i}$ (an unknown fitting parameter), providing only the scale of the overall mobility as $\mu\sim 1/n_\mathrm{i}$.

\begin{figure}[!htb]
\centering
\includegraphics[width=1.0\linewidth]{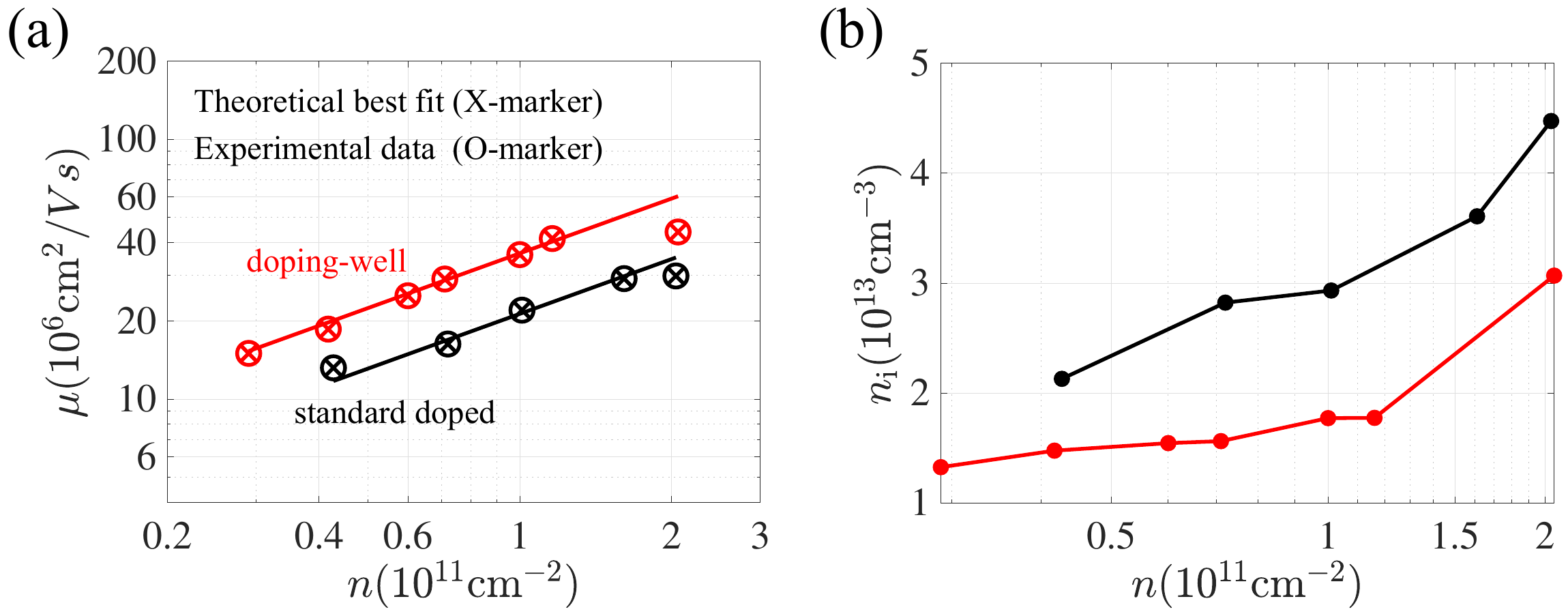}
\caption{(a) Experimental mobility (O-markers) for each sample at a different carrier density given in Fig.~\ref{fig:2}(a) of Ref.~\cite{chungUltrahighqualityTwodimensionalElectron2021} along with the theoretically calculated mobilities (X-markers) that best fit each sample mobility data obtained using the Boltzmann transport theory [Eq.~(\ref{eq:transport_relaxation_time})] with the impurity density $n_\mathrm{i}$ being the tuning parameter. The straight lines represent the power-law relation $\mu\sim n^{0.7}$. The dependence of $\mu$ on $n$ depicted here (and in Ref.~\cite{chungUltrahighqualityTwodimensionalElectron2021}) is not a functional dependence since $\mu$ depends on two independent parameters ($n$ and $n_\mathrm{i}$.) (b) Plot of the background impurity densities extracted in (a) as a function of the carrier density $n$. Here red and black colors indicate with and without doping wells, respectively \cite{chungUltrahighqualityTwodimensionalElectron2021}.
  }
\label{fig:1}
\end{figure}

\begin{figure}[!htb]
\centering
\includegraphics[width=1.0\linewidth]{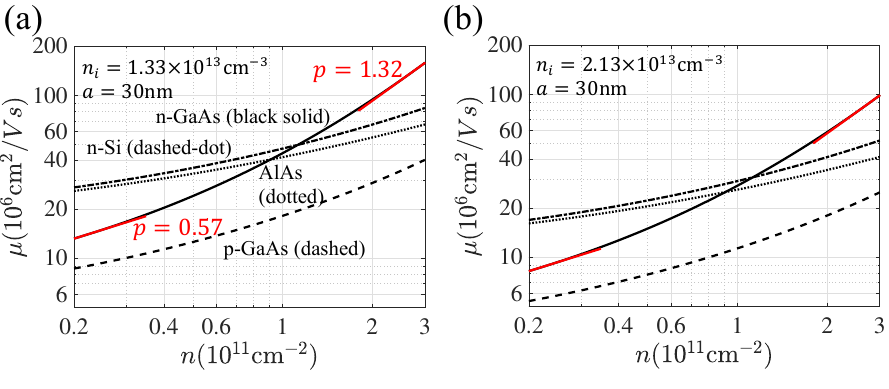}
\caption{Calculated mobility as a function of the carrier density for four different materials using (a) the fixed impurity density corresponding to the lowest mobility in Fig.~\ref{fig:1} and a somewhat higher impurity density of (b) $n_\mathrm{i} =2.13\times10^{13} \mathrm{cm}^{-3}$. The fitted red straight lines in (a) indicate the power-law exponent $p$ (i.e., $\mu\sim n^p$) at low and high densities. For the calculations, we use a fixed value of the quantum well width $a=30\mathrm{nm}$. Note that the varying exponent $p$ happens to be $\sim0.7$ for $n$-GaAs at $n\sim10^{11} \mathrm{cm}^{-2}$ in rough agreement with Fig.~\ref{fig:1}(a).
  }
\label{fig:2}
\end{figure}

In Fig.~\ref{fig:1}(a), we show our calculated mobility in $n$-GaAs, comparing directly with Fig.~\ref{fig:2}(a) of Ref.~\cite{chungUltrahighqualityTwodimensionalElectron2021}. 
By fitting our theory for each sample mobility at the given 2D density (using the experimental sample parameters), we obtain the background impurity density $n_\mathrm{i}$ for each experimental sample as shown in our Fig. \ref{fig:1}(b). Typically $n_\mathrm{i}\sim1-2\times10^{13} \mathrm{cm}^{-3}$ for the carrier density up to $10^{11} \mathrm{cm}^{-2}$, but then it increases to $3\times10^{13} \mathrm{cm}^{-3}$ for $n \sim 2\times10^{11} \mathrm{cm}^{-2}$, seriously degrading the mobility at higher densities ($> 10^{11} \mathrm{cm}^{-2}$), which would have been much higher if the background impurity density could be reduced to $10^{13} \mathrm{cm}^{-3}$. Note that the mobility for the samples without doping wells is much smaller because of higher values of $n_\mathrm{i}$ in these samples as discussed in Ref.~\cite{chungUltrahighqualityTwodimensionalElectron2021}. 
An important message of Fig.~\ref{fig:1} is that the impurity density in the highest mobility sample (so far), $\mu\sim 44\times10^6 \mathrm{cm}^2/Vs$ at $n \sim 2\times10^{11} \mathrm{cm}^{-2}$, is far too high compared with $n_\mathrm{i}$ in the lower mobility (and lower carrier density) samples with $n \sim 10^{11} \mathrm{cm}^{-2}$ in Fig.~\ref{fig:1}. Thus, while the mobility increases with increasing carrier density in Ref.~\cite{chungUltrahighqualityTwodimensionalElectron2021}, it could increase even more if the impurity density could remain the same as at the lower carrier density samples.

By comparing Figs.~\ref{fig:1}(a) and \ref{fig:1}(b), one can find that the carrier mobility increases with increasing impurity density at the lower density regime ($n<10^{11} \mathrm{cm}^{-2}$), which appears to be counterintuitive at first glance. It is important to note, however, that the carrier density also positively correlates with the impurity density as is shown in Fig.~\ref{fig:1}(b). Thus we also need to take into account the effects of increasing carrier density. For two-dimensional electron gas in the weak screening regime, the mobility limited by Coulomb disorders increases linearly with increasing carrier density \cite{dassarmaUniversalDensityScaling2013}. Fig.~\ref{fig:1}(b) shows that the impurity density increases by only a factor of 1.5 or less when the carrier density increases by a factor of more than 3 (roughly from $n=0.3\times10^{11} \mathrm{cm}^{-2}$ to $10^{11} \mathrm{cm}^{-2}$), which explains the counterintuitive behavior of increasing mobility with increasing impurity density.

We show this dramatic effect of impurity scattering in Fig.~\ref{fig:2}(a) by plotting the calculated mobility as a function of 2D carrier density in a fixed $n$-GaAs sample with a fixed impurity density of $n_\mathrm{i}=1.33\times10^{13} \mathrm{cm}^{-3}$, which corresponds to the lowest mobility (and also the lowest carrier density) sample in Fig.~\ref{fig:2} of Ref.~\cite{chungUltrahighqualityTwodimensionalElectron2021} and our Fig.~\ref{fig:1}(a). For this low (but already achieved in a lower density sample) impurity density, the mobility at a 2D carrier density of $n=3\times10^{11} \mathrm{cm}^{-2}$ should be an astronomical $\mu\sim1.5\times10^8 \mathrm{cm}^2/Vs$!  In a real situation, such a high mobility may not be achieved even with an impurity density of $1.33\times10^{13} \mathrm{cm}^{-3}$ and a carrier density of $3\times10^{11} \mathrm{cm}^{-2}$ because various neglected scattering mechanisms such as interface roughness scattering and alloy 
scattering and perhaps even phonon scattering may become operational, but the mobility should still approach $100$ million! We do not know the reason why MBE growth seems to lead to higher impurity density at higher carrier density, but it seems that lowering the impurity density to $<2\times10^{13} \mathrm{cm}^{-3}$ should be feasible given that it appears to be already achieved in lower carrier density samples. In Fig.~\ref{fig:2}(b) we show the predicted mobility as a function of carrier density at a somewhat higher impurity density of $n_\mathrm{i} =2.13\times10^{13} \mathrm{cm}^{-3}$, which is still higher than all our extracted impurity densities in Fig.~\ref{fig:1}(a) except for the highest density sample (where $n_\mathrm{i}=3.1\times10^{13} \mathrm{cm}^{-3}$.) Even at this somewhat elevated impurity density, the 2D $n$-GaAs should approach $100$ million at $n=3\times10^{11} \mathrm{cm}^{-2}$.

In Fig.~\ref{fig:2}, we also show our calculated density-dependent mobility for three other systems:  $p$-GaAs holes in GaAs-AlGaAs quantum wells, $n$-AlAs electrons, and $n$-Si(100) in SiGe quantum wells.  In each case, we assume that the background impurity density is the same as the low numbers achieved in the $n$-GaAs samples-- the results for other values of $n_\mathrm{i}$ can simply be obtained by linear scaling through $\mu \sim 1/n_\mathrm{i}$.  We find that at high enough carrier density, $n > 10^{11} \mathrm{cm}^{-2}$, $n$-GaAs always has the highest mobility, but at lower densities, both $n$-SiGe electrons and $n$-AlAs 2D electrons should have higher mobilities than $n$-GaAs electrons provided, of course, that all systems have equivalent background disorder. But $p$-GaAs 2D holes always have the lowest mobility among the four systems for equivalent disorder. At first glance, our finding of extreme high mobility in lower density $n$-AlAs and $n$-SiGe appears to be incorrect because (1) the current experimental mobility values for both $n$-AlAs and $n$-SiGe 2D electrons are always much lower than that in the 2D $n$-GaAs systems, and (2) the effective mass in both $n$-AlAs ($m \sim 0.5$ in units of electron mass, assuming a transport averaged effective mass incorporating the anisotropy in AlAs) and $n$-SiGe ($m \sim 0.2$ for the 100 Si surface) is much larger than in $n$-GaAs ($m \sim 0.07$), which implies lower mobility intuitively. Our results are, however, correct, and indeed higher effective mass implies a lower effective mobility, as can be seen by the fact that $p$-GaAs 2D holes (with $m \sim 0.4$) in Fig.~\ref{fig:2} always have lower mobility than $n$-GaAs electrons with lighter effective mass. But both AlAs and Si conduction bands have a valley degeneracy of $2$, leading to stronger screening, which makes the low-density mobility limited by Coulomb disorder higher in these systems by virtue of the peculiarity of 2D systems where lower density typically implies stronger effective screening as the dimensionless screening parameter for transport, $q_\mathrm{TF}/2k_\mathrm{F}\sim g_\mathrm{v}\kappa/n^{1/2}$ (where $q_\mathrm{TF}$ and $k_\mathrm{F}$ are the Thomas-Fermi screening and Fermi wavenumbers respectively), increases as $n^{-1/2}$ in 2D with decreasing carrier density. This stronger screening at lower densities leads to the counterintuitive result that at lower densities $n$-SiGe and $n$-AlAs should have higher Coulomb disorder limited mobility than 2D GaAs electrons with no valley degeneracy since screening is proportional to $g_\mathrm{v}$. By contrast, both $p$-GaAs and $n$-GaAs are single valley systems, so the heavier mass GaAs holes always have lower mobility than the lighter GaAs electrons. If the mobility is limited by short-range neutral disorder (e.g., interface roughness or lattice defects), then this phenomenon of higher mobility in $n$-SiGe and $n$-AlAs than in $n$-GaAs would not happen. The question therefore arises why the existing best 2D GaAs holes \cite{watsonExplorationLimitsMobility2012} 
, 2D AlAs electrons \cite{chungMultivalleyTwodimensionalElectron2018}, 
and 2D Si-Ge electrons \cite{huangMobilityEnhancementStrained2012, melnikovUltrahighMobilityTwodimensional2015} 
have much lower mobility than the ones predicted in our theory as shown in Fig.~\ref{fig:2}. For example, the highest reported mobilities in 2D $p$-GaAs, 2D $n$-AlAs, and 2D $n$-SiGe quantum wells are, respectively, $2.3\times10^6 \mathrm{cm}^2/Vs$ at $n= 6.5\times10^{10} \mathrm{cm}^{-2}$, $2.4\times10^6 \mathrm{cm}^2/Vs$ at $n=2.2\times 10^{11} \mathrm{cm}^{-2}$, and $2.4\times10^6 \mathrm{cm}^2/Vs$ at $n=10^{11} \mathrm{cm}^{-2}$. This indicates background charged impurity densities of $8\times10^{13} \mathrm{cm}^{-3}$ ($p$-GaAs), $3\times10^{14} \mathrm{cm}^{-3}$ ($n$-AlAs) and $2\times10^{14} \mathrm{cm}^{-3}$ ($n$-SiGe) respectively, as compared with the results in our Fig.~\ref{fig:2} (and appropriately linearly scaled by the impurity density.) We therefore conclude that $p$-GaAs, $n$-AlAs and $n$-SiGe samples are still much lower quality than the state of the arts $n$-GaAs samples reported in Ref.~\cite{chungUltrahighqualityTwodimensionalElectron2021} (see Table.~\ref{table:1} for a summary.) 

\begin{table} \caption{Highest reported experimental mobilities($\mu_\mathrm{peak}$) and their corresponding carrier densities($n$). $n_\mathrm{i}$ indicates estimated background impurity densities obtained using Fig.~\ref{fig:2}, which are much larger than that of the state of the arts $n$-GaAs samples reported in Ref.~\cite{chungUltrahighqualityTwodimensionalElectron2021}. } 
    \begin{ruledtabular} 
        \begin{tabular}{c c c c} 
        Material & $\mu_\mathrm{peak}$($\mathrm{cm}^2/Vs$) & $n(\mathrm{cm}^{-2})$ & $n_\mathrm{i}(\mathrm{cm}^{-3})$\\ 
        \hline
        $p$-GaAs    & $2.3\times 10^6$ & $6.5\times10^{10}$ & $8\times10^{13}$\\
        $n$-AlAs    & $2.4\times 10^6$ & $2.2\times10^{11}$ & $3\times10^{14}$\\ 
        $n$-SiGe    & $2.4\times 10^6$ & $1.0\times10^{11}$ & $2\times10^{14}$\\ 
        \end{tabular} 
    \end{ruledtabular} 
    \label{table:1}
\end{table}

\begin{figure}[!htb]
\centering
\includegraphics[width=1.0\linewidth]{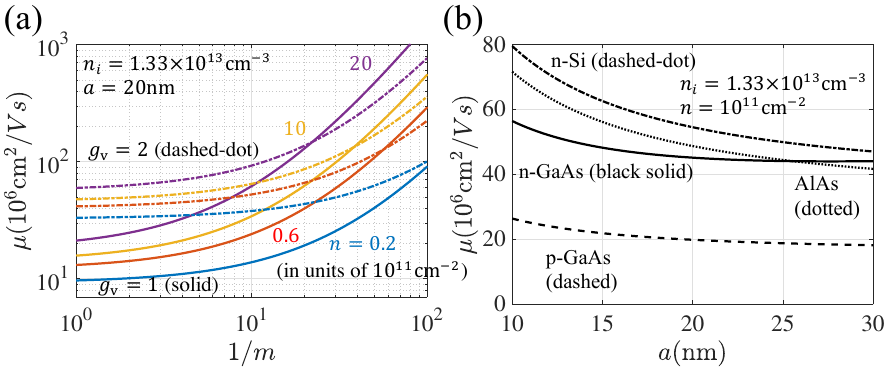}
\caption{(a) Calculated mobilities as a function of the inverse of the effective mass at different carrier densities $n$ with the valley degeneracy of $g_v=1$ and $2$, showing the mobility dependence on both valley degeneracy and effective mass. Here $m$ is in units of electron mass. (b) Calculated mobilities as a function of the quantum well width $a$ for four different materials at a fixed carrier density of $n=10^{11}\mathrm{cm}^{-2}$, showing the mobility dependence on the well width. For both results in (a) and (b), we use the lowest best-fit impurity density of the $n$-GaAs sample with a doping-well ($n_\mathrm{i}=1.33\times 10^{13}\mathrm{cm}^{-3}$), which we obtain in Fig.~\ref{fig:1}. 
  }
\label{fig:3}
\end{figure}

\begin{figure}[!htb]
\centering
\includegraphics[width=1.0\linewidth]{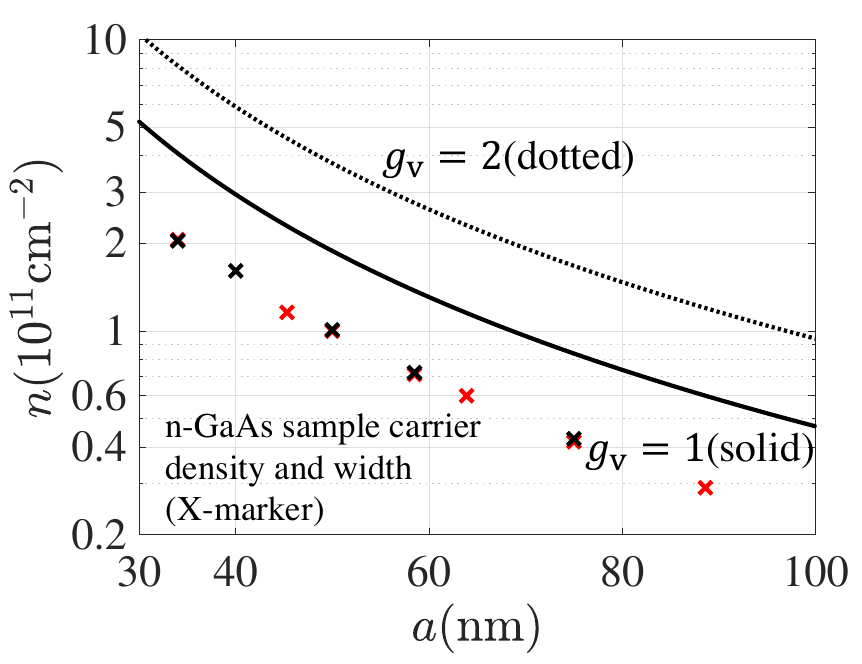}
\caption{Phase diagram showing the regime where the single subband approximation is valid. The approximation is valid when the carrier density (at a given well width $a$) is below the the plotted curves, which represent the onset density where the second subband occupation occurs. The X-markers indicate the carrier densities (and the corresponding quantum well widths) of the high mobility $n$-GaAs samples in Ref.~\cite{chungUltrahighqualityTwodimensionalElectron2021}, which are used in our calculations of Fig.~\ref{fig:1}.}
\label{fig:4}
\end{figure}

In Fig.~\ref{fig:3}, we show the dependence of the mobility on the well width, effective mass and valley degeneracy, clearly demonstrating the asymptotic behavior of $\mu \sim m^0$ for large mass and $\mu\sim m^{-2}$ for small mass. These asymptotic dependences follow from the screening behavior of the Coulomb disorder with large (small) mass corresponding to strong (weak) screening limits. Increasing $g_v$ enhances screening, and thus increases mobility if the other parameters remain fixed, explaining why both $n$-AlAs and $n$-SiGe have higher mobilities than $n$-GaAs at the low-density strong screening limit. The well-width dependence of the mobility is rather modest, but we warn that if the width is too small (large), interface (intersubband) scattering may become important.  In Fig.~\ref{fig:4}, we provide a `phase diagram' for the regime where the single subband approximation used in our theory is valid for a given well width and carrier density. As is obvious from Fig.~\ref{fig:4}, all our results in Figs.~\ref{fig:1}-\ref{fig:3} are in the 1-subband occupancy regime as are all the experimental results with which we compare.

\section{Discussion and Conclusion}
Our main conclusions are : (1) the recent breakthrough in $n$-GaAs 2D MBE growth has led to unprecedented low (high) background disorder (mobility), but the current ultrahigh mobility of 44 million is not the optimal mobility at the $10^{11} \mathrm{cm}^{-2}$ carrier density since lower disorder samples have already been studied in the earlier literature-- therefore, further mobility improvement in the near future is likely; (2) compared with the 2D $n$-GaAs samples of ultrahigh mobility, other MBE-grown modulatio$n$-doped 2D carrier systems (e.g., $p$-GaAs, $n$-AlAs, $n$-SiGe) are still very dirty, and substantial improvement in their mobilities, even surpassing the $n$-GaAs mobility at low densities, should be possible in the future with improvement in the growth quality with fewer background impurities. We emphasize that our Boltzmann transport theory calculation is essentially exact for all the results shown in this work because the calculated mobility satisfies the condition $k_\mathrm{F}L\gg1$ (with $L$ being the transport mean free path) with the estimated $k_\mathrm{F}L$ values being in the range of $10^3$-$10^5$ in our calculated results.

We note that the experimental mobility results in Fig.~\ref{fig:1}(a) scale with density according to the empirical relation $\mu\sim n^p$, with $p\sim 0.7$. This is, however, merely a coincidence and not a fundamental functional relationship since each experimental data point in Fig.~\ref{fig:1}(a) represents a different sample with varying carrier density and varying background impurity density. For fixed background disorder, as in our theoretical results in Fig.~\ref{fig:2}(a), there is no strict scaling with the exponent $p$ varying slowly with density $n$, and $p(n)$ increasing with increasing $n$ (and also being somewhat dependent on the material.) The physics here is screening-- low (high) density screens the background Coulomb disorder strongly (weakly) with $p(n)$ tending toward 1/2 (3/2) as $q_\mathrm{TF}/2k_\mathrm{F}$ tends toward infinity (zero) \cite{dassarmaUniversalDensityScaling2013} 
(For pure remote scattering by the modulation layer dopants, the exponent $p$ is always 3/2 except for very low carrier densities.)
For $n\sim10^{11} \mathrm{cm}^{-2}$, $n$-GaAs 2D systems manifests $p \sim 0.7$, but it is by no means a constant exponent.  This is shown in our Fig.~\ref{fig:1} by straight lines indicating the effective exponent $p$ at low and high densities.

Finally, we discuss one immediate implication of the ultra-high mobility achieved in Ref.~\cite{chungUltrahighqualityTwodimensionalElectron2021}. The reported energy gap for the $5/2$ FQHE $\sim 0.82$K at $n\sim 10^{11} \mathrm{cm}^{-2}$ is by far the highest activation gap ever reported for this no$n$-Abelian FQH state at any density, the previous record being a gap of $0.54$K at $n\sim 3.2\times10^{11} \mathrm{cm}^{-2}$ \cite{choiActivationGapsFractional2008} and $0.6$K at $3.4\times10^{11}\mathrm{cm}^{-2}$ \cite{qianQuantumLifetimeUltrahigh2017}.
Converting both gaps into dimensionless Coulomb energy units and incorporating the finite width correction to the Coulomb energy \cite{zhangExcitationGapFractional1986}, 
we find that the current $5/2$ experimental gap in Ref.~\cite{chungUltrahighqualityTwodimensionalElectron2021} is $\sim 70$\text{\%}
of the theoretically estimated ideal $5/2$ FQHE gap \cite{morfExcitationGapsFractional2002}
whereas the earlier highest measured gaps are roughly 40\text{\%} of the ideal theoretical gap.  This 30\text{\%} improvement in the measured effective gap for the 5/2 FQHE is a significant advance, which should lead to a rethinking of the role of the 5/2 no$n$-Abelian FQHE as a platform for topological quantum computation \cite{dassarmaTopologicallyProtectedQubits2005}
since the current measured topological FQHE gap of 0.82K is already higher than that estimated topological gap ($\sim0.6$K) in the semiconductor nanowire platform which is actively being studied for topological Majorana qubits \cite{sauNonAbelianQuantumOrder2010}. 
In fact, incorporating the Landau level mixing effect approximately \cite{morfDisorderFractionalQuantum2003}, 
the measured gap \cite{chungUltrahighqualityTwodimensionalElectron2021} may be approaching 90\text{\%} of the ideal theoretically expected 5/2 FQHE energy gap.

\section{Acknowledgement} \label{sec:acknowledgement}
This work is supported by the Laboratory for Physical Sciences.

%


\end{document}